\documentclass[10pt, twocolumn]{IEEEtran}
\normalsize
\usepackage{amsthm,amssymb,amsmath,tikz,graphics,cite,float,graphicx,epstopdf,epsfig,verbatim,url}
\usetikzlibrary{automata}
\usetikzlibrary{shapes,arrows}

\usepackage[utf8]{inputenc}

\usepackage{capt-of}
\usepackage[noend]{algpseudocode}
\usepackage{algorithmicx}
\usepackage[ruled]{algorithm}

\usepackage{lipsum}

\usepackage{mdwtab}
\usepackage{subfigure}
\usepackage{placeins}
\usepackage{setspace}

\allowdisplaybreaks

\theoremstyle{plain}
\theoremstyle{remark}

\begin{document}
\newtheorem{theorem}{Theorem}
\newtheorem{lemma}{Lemma}
\newtheorem{conjecture}{Conjecture}
\newtheorem{corollary}{Corollary}
\newtheorem{definition}{Definition}
\newtheorem{property}{Property}
\newtheorem{remark}{Remark}

%\lipsum
\bibliographystyle{plain}

\title{On the Power Allocation Limits for Downlink Multi-user NOMA with QoS}

\author{\IEEEauthorblockN{Jos\'{e} Armando Oviedo and  Hamid R. Sadjadpour}\\
\IEEEauthorblockA{Department of Electrical Engineering, University of California, Santa Cruz\\
Email: \{xmando, hamid\}@soe.ucsc.edu}}

\maketitle

\IEEEpeerreviewmaketitle

\begin{abstract}
	The fundamental power allocation requirements for NOMA systems with minimum quality of service (QoS) requirements are investigated. For any minimum QoS rate $R_0$, the limits on the power allocation coefficients for each user are derived, such that any power allocation coefficient outside of these limits creates an outage with probability equal to 1. The power allocation coefficients that facilitate each user's success of performing successive interference cancellation (SIC) and decoding its own signal are derived, and are found to depend only on the target rate $R_0$ and the number of total users $K$. It is then proven that using these power allocation coefficients create the same outage event as if using orthogonal multiple access (OMA), which proves that the outage performance of NOMA with a fixed-power scheme can match that of OMA for all users simultaneously. Simulations confirm the theoretical results, and also demonstrate that a power allocation strategy exists that can improve the outage performance of NOMA over OMA, even with a fixed-power strategy.

\end{abstract}

\section{Introduction}
Non-orthogonal multiple access (NOMA) is a multiple access approach that will be implemented into future wireless standards to increase spectral efficiency and help future systems achieve the throughput demands of the near future \cite{3GPP:MUST}. NOMA takes advantage of the power-domain by employing superposition coding (SC) at the transmitter and successive interference cancellation (SIC) at the receiver serving multiple users simultaneously. In the downlink, the base-station (BS) transmits the signals to multiple users over an entire transmission period and bandwidth. Each user then obtains its own message by employing SIC to decode and remove the interference of signals with greater power and lower data-rates than its own. This is in contrast to orthogonal multiple access (OMA), which assigns a non-overlapping time or frequency slot to each user in order to eliminate interference. However, since achievable date rates are restricted more by time/bandwidth than by power, NOMA outperforms OMA in terms of achieving higher data rates \cite{InfTh:CT}. 

Although it is proven in \cite{FairNOMAFull} that there always exists a power allocation approach for NOMA that can outperform OMA for the general multi-user NOMA case in terms of information capacity, this power allocation strategy relies on having perfect channel state information at the transmitter, which is not a realistic assumption in practical systems. Furthermore, given that wireless systems typically require that users be able to sustain a minimum rate, it is important to demonstrate that NOMA systems are capable of facilitating users to obtain a channel that can support the minimum rate at least as well as OMA.

The purpose of this work is to demonstrate that for any minimum QoS rate $R_0$ required by the system, there exists a fundamental power allocation coefficient set that makes the probability non-zero of experiencing an outage. The rest of this paper is organized as follows: section \ref{sec:previous} outlines the important previous work on NOMA with minimum QoS rates, section \ref{sec:system} describes the downlink wireless system parameters, section \ref{sec:QoS} then delves into the theoretical details of NOMA systems with a minimum QoS requirement, section \ref{sec:results} demonstrates the validity of the theoretical results, and section \ref{sec:conclusion} outlines the conclusions and suggestions for continuing the work.

\section{Previous Work on NOMA} \label{sec:previous}

The outage probability of NOMA was first investigated in \cite{NOMA-HARQ:Choi}, where multiple users transmit simultaneously to multiple receivers, and it is shown the outage probability is improved when NOMA is combined with H-ARQ vs OMA with H-ARQ. However, this work uses a uniformly flat power allocation approach. The authors in \cite{NOMA-RDU:DYFP} are the first work to show that the power allocation and interference coefficients of each user are fundamentally dependent on the particular user's required rate, and thus the wrong selection of coefficients can lead to an outage with probability equal to 1. The usage of NOMA in a cogintive-radio inspired approach was mentioned in \cite{5GNOMA:DFP}, where a user with weak channel condition is seen as the primary user and is provided as much power as needed in order to achieve its minimum rate, and the user with stronger channel is treated as the secondary user and receives any remaining power not allocated to the weaker user, and the outage probability of both is shown to clearly depend on pairing users with stronger channels. 

A couple of works have focused on utilizing the rate achieved using OMA as the minimum rate required by a NOMA user, leading to a focus on finding the power allocation coefficients that achieve this condition, and demonstrate the outage performance for NOMA outperforms the outage performance of OMA. The region of power allocation coefficients that allow NOMA to outperform OMA in the downlink is first defined for the two-user case in \cite{FairNOMA:Infocom}. The authors in \cite{DNOMA:YDFA} then use a power allocation approach in this region to analyze the outage performance and diversity orders of two paired users, according to their relative channel gains, and extend the work to the uplink case. In \cite{FairNOMAFull}, the power allocation coefficients for a multi-user NOMA system that always outperforms OMA are proven to always sum to less than or equal to 1, and hence show that a valid power allocation strategy for NOMA always exists that outperforms OMA in terms of capacity, while using less power than OMA.

The main contribution of this work is to expand on the fundamental power allocation requirements of NOMA system that support more than two simultaneous users, by focusing on systems that require a minimum QoS rate. This is achieved by doing the following:
\begin{itemize}
	\item The fundamental limits for the power allocation coefficients are derived such that no users have an outage probability of 1. The limits are found to depend only on the minimum rate required and the total number of users, and not on the channel gains or the transmit SNR.
	\item A power allocation strategy is then derived that is proven to allow each user to perform SIC on signals for users with lesser channel gains, and then decode its own signal, all with equal outage events. This is important because selecting power allocation coefficients while assuming SIC to be successful can lead to the event where an outage encountered during SIC causes outages on all subsequent rates to decode. 
	\item Since the power allocation strategy derived depends only on the minimum rate and and number of users, it is inherently a fixed-power strategy. However, it is proven that this strategy provides NOMA with the same outage event as OMA. 
\end{itemize}

\section{System Model and Capacity}\label{sec:system}
Consider a wireless downlink system with a base-station that has $M$ transmit antennas, and $K$ user equipments that each possess $N$ antennas. The base-station will transmit $K$ signals, each containing the information for one of the $K$ users. Let the signal for user $n$ be $x_n$, $n=1,\ldots, K$, $\mathbb{E}[|x_n|]=1$. The $K$ signals are all transmitted with transmit SNR $\xi$, along the same transmit beamforming vector $\boldsymbol{v}$. The vector $\boldsymbol{v}$ ($\|\boldsymbol{v}\|^2=1$) is chosen to be chosen randomly from an isotropic distribution. The wireless channel matrix from the $M$ transmit antennas to the $N$ receive antennas of each user $n$ is $\boldsymbol{H}_n\in\mathbb{C}^{N\times M}$. At the receiver $n$, the received combining vector $\boldsymbol{u}_n$ is selected using maximum ratio combining to optimize the received signal power ($\|\boldsymbol{u}_n\|^2=1$). Lets define the composite channel gain as $G_n = \boldsymbol{u}_n^\mathsf{H}\boldsymbol{H}_n\boldsymbol{v}$. The channel gains of users are ordered as $|G_1|^2<\cdots<|G_K|^2$. 

In the case of OMA, the received signal at user $n$ is given by
\begin{equation}
	y_n = \sqrt{\xi}G_n x_n + z_n',
\end{equation}
where $z_n'=\boldsymbol{u}_n^\mathsf{H}\boldsymbol{z}_n$, and $\boldsymbol{z}_n\sim\mathcal{CN}(0,\mathbf{I}_K)$. Since user $n$ is given $\frac{1}{K}$ of the time/frequency for its orthogonal slot, the capacity of user $n$ using OMA is then given by
\begin{equation}
	C_{n}^\mathrm{oma} = \frac{1}{K}\log_2\left(1 + \xi |G_n|^2 \right).
\end{equation}

For the NOMA system, user $n$ has power allocation coefficient $a_n$, such that $\sum_{n=1}^K a_n = 1$. The received signal at user $n$ is
\begin{equation}
	r_n = G_n\sum_{l=1}^K x_l\sqrt{a_l\xi} + z_n',
\end{equation}
Using successive interference cancellation, the receiver at user $n$ will decode the messages of users $m<n$ in ascending order, starting with $m=1$. Therefore, user $n$ will perform SIC on the signals of user $m=1,\ldots,n-1$, which have the form
\begin{equation}
    y_{n\rightarrow m} = G_n(\underbrace{x_m\sqrt{a_m\xi}}_\text{user-$m$ signal for SIC}+\sum_{l=m+1}^K x_l\sqrt{a_l\xi}) + z_n, 
\end{equation}
until it can obtain the intended signal at user $m$ is given by 
\begin{equation}
	y_n = G_n(\underbrace{x_n\sqrt{a_n\xi}}_\text{user-$n$ signal}+\sum_{l=n+1}^K x_l\sqrt{a_l\xi}) + z_n,
\end{equation}
where $\sum_{l=n+1}^K a_lx_l$ are the signals that are sent with messages encoded at rates beyond the capacity region of user $n$, and thus are treated as interference.

Given the order of the channel gains, each user $n$ will decode the messages intended for each user $m<n$ before decoding its own message. Let the power allocation coefficients be $a_1,\ldots,a_K$, then each user $n<K$ has capacity 
\begin{equation}
	C_n(a_1,\ldots,a_K)=\log_2\left(1 + \frac{a_n\xi |G_n|^2}{1+\xi |G_n|^2\sum_{l=n+1}^K a_l} \right),
\end{equation}
and user $K$ has capacity 
\begin{equation}
	C_K(a_1,\ldots,a_K)=\log_2\left(1 + a_K\xi |G_K|^2 \right).
\end{equation}

Meanwhile, for each user $n$ to achieve its capacity, it must have the capacity to decode the messages sent to all users $m<n$, and subtract their signals from the composite signal received. The capacity for user $n$ to decode user $m$'s message is given by 
\begin{equation}
	C_{n\rightarrow m}(a_1,\ldots,a_K) = \log_2\left(1 + \frac{a_m\xi |G_n|^2}{1+\xi |G_n|^2\sum_{l=m+1}^K a_l} \right).
\end{equation}

\section{NOMA power allocation for minimum QoS rate systems} \label{sec:QoS}
Suppose that the system requires that each user scheduled satisfy a minimum QoS rate requirement $R$. This means that all users must have capacity greater than a rate $R_0$. In other words, for each user $n$, it must be true that 
\begin{equation}
	C_n(a_1,\ldots,a_K)>R_0\hspace{2mm}\text{ AND }\hspace{2mm}C_{n\rightarrow m}(a_1,\ldots,a_K)>R_0.
\end{equation}
Define the events $\mathcal{B}_n = \{C_n(a_1,\ldots,a_K)>R_0\}$ and $\mathcal{B}_{n\rightarrow m}=\{C_{n\rightarrow m}(a_1,\ldots,a_K)>R_0\}$. We now arrive at a fundamental fact for power allocation for NOMA systems with minimum QoS rate $R_0$.
\begin{theorem}\label{thm:power_limit}
    For a $K$-user NOMA systems that requires that all $K$ users have capacity greater than a minimum QoS rate $R_0$, the interference received by the signal carrying user $n$'s message has total power coefficient $A_n=\sum_{l=n+1}^K a_l, \forall n=1,\ldots,K$. If $\exists n$ such that $A_n > 2^{-R_0n}$, then $\forall l\geq n$,
    \begin{equation}
        \mathrm{Pr}\{\mathcal{B}_l\} = \mathrm{Pr}\{\mathcal{B}_{l\rightarrow n}\} = 0,
    \end{equation} 
    and thus SIC will fail. 
\end{theorem}
\begin{IEEEproof}
	For any specific user $n$, suppose that $A_{n-1} < 2^{-R_0(n-1)}$ and $A_n > 2^{-R_0n}$. Since $A_{n-1}=a_n+A_n$, it follows that 
	\begin{align}
		&a_n+A_n < 2^{-R_0(n-1)} \nonumber \\
		&\Longrightarrow a_n < 2^{-R_0(n-1)}-A_n < 2^{-R_0(n-1)} - 2^{-R_0n},
	\end{align}
	so $a_n < 2^{-R_0(n-1)} - 2^{-R_0n}$. The events $\mathcal{B}_n$ and $\mathcal{B}_{l\rightarrow n}$ can be written in the form
	\begin{align}
		&\log_2\left( 1 + \frac{a_n\xi |G_l|^2}{A_n\xi |G_l|^2+1}\right) > R_0, l\geq n \nonumber\\
		\Rightarrow&\xi |G_l|^2 ( a_n -(2^{-R_0}-1)A_n) > 2^{R_0}-1.
	\end{align}
	Since $a_n<2^{-R_0(n-1)} - 2^{-R_0n}$ and $A_n>2^{-R_0n}$, 
	\begin{align}
		&a_n -(2^{R_0}-1)A_n \nonumber\\
		&<2^{-R_0(n-1)} - 2^{-R_0n} -(2^{R_0}-1)A_n \nonumber\\
		&<2^{-R_0(n-1)} - 2^{-R_0n} -(2^{R_0}-1)2^{-R_0n} \nonumber\\
		&=0.
	\end{align}
	This leads to 
	\begin{align}
		&\xi |G_l|^2 ( a_n -(2^{-R_0}-1)A_n) > 2^{R_0}-1 \nonumber\\
		\Rightarrow& |G_l|^2 < \frac{2^{-R_0}-1}{\xi(a_n-(2^{-R_0}-1)A_n)} < 0,
	\end{align}
	which is an impossible event since $|G_n|^2 > 0, \forall n=1,\ldots,K$. Hence, if $A_n>2^{-R_0(n-1)}$, then $ \mathrm{Pr}\{\mathcal{B}_n\} = \mathrm{Pr}\{\mathcal{B}_{l\rightarrow n}\} = 0$. \\
	
	Now suppose that $A_n > 2^{-R_0n}, \forall n$, then it must be true that $A_1 > 2^{-R_0}$. This will avoid the previous impossible event. However, if this is true, then rearranging events $\mathcal{B}_1$ and $\mathcal{B}_{l\rightarrow 1}$ gives rise to the inequality
	\begin{align}
		&\xi |G_l|^2 (a_1 -(2^{R_0}-1)A_1) > 2^{R_0}-1,
	\end{align}
	where the value inside the parentheses must be greater than zero. Therefore, 
	\begin{align}
		&0< a_1 -(2^{R_0}-1)A_1 < a_1 -(2^{R_0}-1)2^{-R_0} \nonumber \\
		\Rightarrow& a_1 > 1-2^{-R_0}. 
	\end{align}
	However, since $a_1+A_1=1$, then
	\begin{align}
		&1 = a_1+A_1 > (1-2^{-R_0}) + 2^{-R_0} = 1.  
	\end{align}
	This is a contradiction. Therefore if $A_n>2^{-R_0n}, \forall n$, then having $\mathrm{Pr}\{\mathcal{B}_n\} >0$ and $\mathrm{Pr}\{\mathcal{B}_{l\rightarrow n}\}>0$ requires $a_1+A_1 = \sum_{n=1}^K a_n> 1$, which is not possible.

	Hence, for any user $n$ with $A_n>2^{-R_0n}$, $\mathrm{Pr}\{\mathcal{B}_n\} = \mathrm{Pr}\{\mathcal{B}_{l\rightarrow n}\} = 0$.	
\end{IEEEproof}
Theorem \ref{thm:power_limit} demonstrates that for any minimum rate $R_0$, there is a fundamental set from which the power allocation coefficients must come from. It also demonstrates clearly that as $R_0$ increases, the values of $a_n$ decrease rapidly, which means that if a user's channel is too weak, the power needed to avoid an outage may become prohibitively too large. 

Note that this does not indicate that the rate for user $n$ is guaranteed if $A_n<2^{-R_0n}$, since the total power available for allocation to users $n,\ldots,K$ may be less than $2^{-R_0n}$ to begin with, yet a value for $A_n<2^{-R_0n}$ can still make an outage certain, as outlined in \cite{NOMA-RDU:DYFP}.

Given that $(a_1,\ldots, a_K)$ come from the set such that $A_n<2^{-R_0n}, \forall n<K$ (user-$K$'s signal does not receive interference, so long as it decodes all other signals successfully), the events $\mathcal{B}_n$ and $\mathcal{B}_{n\rightarrow m}, n>m$, can be rewritten as 
\begin{align}
	&\mathcal{B}_n = \left\{ |G_n|^2>\frac{2^{R_0}-1}{\xi(a_n+(1-2^{R_0})\sum_{l=n+1}^K a_l)} \right\}
\end{align}
and 
\begin{align}
	&\mathcal{B}_{n\rightarrow m} = \left\{ |G_n|^2>\frac{2^{R_0}-1}{\xi(a_m+(1-2^{R_0})\sum_{l=m+1}^K a_l)} \right\}.
\end{align}
Therefore, the signal for user-$m$ $y_m$ can only be decoded if the channel gains for all users $n>m$ have channel gains
\begin{equation}
    |G_n|^2>\frac{2^{R_0}-1}{\xi(a_m - (2^{R_0}-1)\sum_{l=m+1}^K a_l)}, n\geq m.
\end{equation}
For the case of user-$K$, the event
\begin{align}\label{eq:B_K}
	\mathcal{B}_K = \{\log_2(1+a_K\xi|G_K|^2)>R_0\} = \left\{|G_K|^2>\frac{2^{R_0}-1}{a_K\xi}\right\}
\end{align}

	A user $n$ can only successfully decode its own message if all of the events $\mathcal{B}_{n\rightarrow m}$ are true for $m<n$. Otherwise, if any of these events fail to happen, then $\mathcal{B}_n$ has a zero probability of occuring since the capacity $C_n(a_1,\ldots,a_K)$ is no longer achievable. One way to solve this problem is that all events $\mathcal{B}_n$ and $\mathcal{B}_{n\rightarrow m}, \forall m<n$, should have the same probability of occuring for each user $n$, i.e. they should become the same event. This leads to the following result.
\begin{lemma}\label{lem:ampower}
    For a $K$-user NOMA system with QoS minimum rate $R_0$ and power allocation coefficients $\{a_1, \ldots, a_K\}$ such that $\sum_{n=1}^K a_n=1$, if $\forall n=1,\ldots,K$,
    \begin{align}
    	\mathcal{B}_n = \mathcal{B}_{n\rightarrow m}, \forall m<n, 
    \end{align}
    then 
    \begin{equation}
		a_n = \frac{2^{(K-n)R_0}(2^{R_0}-1)}{2^{KR_0}-1}, n=1,\ldots, K.
    \end{equation}
\end{lemma}
\begin{IEEEproof}
Note that for each particular $m<n$, $\mathcal{B}_m$ and $\mathcal{B}_{n\rightarrow m}$ have on the right side of the inequality the same exact expression 
\begin{equation}
	\frac{2^{R_0}-1}{\xi(a_m-(2^{R_0}-1)\sum_{l=m+1}^K a_l) }. 
\end{equation}
Therefore, in equating the expressions for $m=1,\ldots,n$, for each user $n=2,\ldots,K$ (user $1$ does not perform SIC), it is exactly the same as setting up the following system of $K-2$ equations for $n=2,\ldots,K-1$,
\begin{align}
	&\frac{2^{R_0}-1}{\xi(a_{n-1}+(1-2^{R_0})\displaystyle\sum_{l=n}^K a_l)}=\frac{2^{R_0}-1}{\xi(a_n+(1-2^{R_0})\displaystyle\sum_{l=n+1}^K a_l)}, \nonumber\\
	\label{eq:genoutset}&\Rightarrow a_{n-1} = a_n 2^{R_0}, n=2,\ldots,K-1.
\end{align}
Using event $\mathcal{B}_K$ gives
\begin{align}
	&\frac{2^{R_0}-1}{\xi(a_{K-1}+(1-2^{R_0})a_K)}=\frac{2^{R_0}-1}{a_K\xi} \nonumber\\
	\label{eq:maxoutset}\Rightarrow&a_{K-1} = a_{K}2^{R_0}.
\end{align}
Combining the results from equations (\ref{eq:genoutset}) and (\ref{eq:maxoutset}) gives
\begin{align}
	\label{eq:amresult}\Rightarrow& a_n = a_K 2^{(K-n)R_0}, n=1,\ldots,K.
\end{align}
Using the fact that $1 = \sum_{n=1}^K a_n$, and substituting $l=K-n$, the power allocation coefficients can be substituted with equation \eqref{eq:amresult} to yield
\begin{align}
	&1 = a_K\sum_{n=1}^{K} 2^{(K-n)R_0}=a_K\sum_{l=0}^{K-1} 2^{lR_0}=a_K\frac{2^{KR_0}-1}{2^{R_0}-1}\\
	&\Rightarrow a_K = \frac{2^{R_0}-1}{2^{KR_0}-1}
\end{align}
Thus, the power allocation coefficient for user $l$ is 
\begin{equation}
	a_n = \frac{2^{(K-n)R_0}(2^{R_0}-1)}{2^{KR_0}-1}, n=1,\ldots, K.
\end{equation}
\end{IEEEproof}
With lemma \ref{lem:ampower}, it is shown that a power allocation set always exists such that if a user $n$ can decode its own information, it can decode all of the information of users $m<n$, because the events become the same condition that the channel gain $G_n$ must overcome. 

Using the power allocation proven in lemma \ref{lem:ampower}, the following result is derived regarding the outage probabilities of the users using fixed-power allocation.
\begin{theorem}\label{thm:nomaomapower}
	If 
	\begin{equation}
		a_n = \frac{2^{(K-n)R_0}(2^{R_0}-1)}{2^{KR_0}-1}, n=1,\ldots, K,
	\end{equation} 
	then $\forall n, \mathcal{B}_n\ = \mathcal{B}_{n\rightarrow m} = \mathcal{B}_n^\text{oma}$, where $\mathcal{B}_m^\text{oma} = \{\frac{1}{K}\log_2(1+\xi |G_n|^2)>R_0\}$.
\end{theorem}
\begin{IEEEproof}
	For the case of user $K$, it is easily shown that 
	\begin{align}
		&\mathcal{B}_K = \{\log_2(1 + a_K\xi |G_K|^2) > R_0\} \\
		& = \left\{ \log_2\left(1 + \frac{2^{R_0}-1}{2^{KR_0}-1}\xi |G_K|^2\right)> R_0\right\} \nonumber\\
		& = \left\{ \frac{2^{R_0}-1}{2^{KR_0}-1}\xi |G_K|^2> 2^{R_0}-1\right\} \nonumber\\
		& = \left\{ \xi |G_K|^2 > 2^{KR_0}-1\right\} \nonumber\\
		& = \{\tfrac{1}{K}\log_2(1+\xi |G_K|^2)>R_0\} = \mathcal{B}_K^\text{oma}.
	\end{align}
	For the general case of user $n$, it follows that
	\begin{align}
		&\mathcal{B}_n = \left\{ \log_2\left(1 + \frac{a_n\xi |G_n|^2}{1+ \xi |G_n|^2\sum_{l=n+1}^K a_l} \right) > R_0\right\}\\
		&= \left\{ \log_2\left(1 + \frac{\frac{2^{R_0(K-n)}(2^{R_0}-1)}{2^{KR_0}-1}\xi |G_n|^2}{1+ \xi |G_n|^2\frac{2^{R_0}-1}{2^{KR_0}-1}\displaystyle\sum_{l=n+1}^K \hspace{-2mm}2^{R_0(K-l)}} \right) > R_0 \right\} \nonumber\\
		&= \left\{ \log_2\left(1 + \frac{\frac{2^{R_0(K-n)}(2^{R_0}-1)}{2^{KR_0}-1}\xi |G_n|^2}{1+ \xi |G_n|^2\frac{2^{R_0}-1}{2^{KR_0}-1}\cdot\frac{2^{R_0(K-n)}-1}{2^{R_0}-1}} \right) > R_0 \right\} \nonumber\\
		&= \left\{ \log_2\left(1 + \frac{\frac{2^{R_0(K-n)}(2^{R_0}-1)}{2^{KR_0}-1}\xi |G_n|^2}{1+ \xi |G_n|^2\frac{2^{R_0(K-n)}-1}{2^{KR_0}-1}  }\right) > R_0 \right\} \nonumber\\
		&= \left\{  \frac{\frac{2^{R_0(K-n)}(2^{R_0}-1)}{2^{KR_0}-1}\xi |G_n|^2}{1+ \xi |G_n|^2\frac{2^{R_0(K-n)}-1}{2^{KR_0}-1}  } > 2^{R_0}-1 \right\} \nonumber\\
		&= \left\{\frac{2^{R_0(K-n)}}{2^{KR_0}-1}\xi |G_n|^2 > 1+ \xi |G_n|^2\frac{2^{R_0(K-n)}-1}{2^{KR_0}-1} \right\} \nonumber\\
		&= \left\{ \frac{\xi |G_n|^2}{2^{KR_0}-1}>1 \right\} \nonumber\\
		&= \left\{ \xi |G_n|^2>2^{KR_0}-1\right\} \nonumber\\
		& = \{\tfrac{1}{K}\log_2(1+\xi |G_n|^2)>R_0\}=\mathcal{B}_n^\text{oma}.
	\end{align}
\end{IEEEproof}
This theorem is stating that the outage performance of NOMA is always equal to OMA for every single user so long as the power allocation given by lemma \ref{lem:ampower} is used. Note that this does not imply that the capacity of user-$n$'s channel using NOMA with $a_n$ given by lemma \ref{lem:ampower} is equal to the capacity of user-$n$'s channel using OMA. 

This result is powerful because it can set an initial point for search algorithms to optimize the power allocation coefficients in order to minimize the overall outage probability of the network, where the initial points puts the system to perform at least as good as OMA. Therefore, this suggests that a strategy can always be found where optimum overall outage probability is better than when using OMA, even when using fixed-power.

\section{Comparison of theoretical and simulation results} \label{sec:results}
\begin{figure}
	\centering
	\includegraphics[scale=0.65]{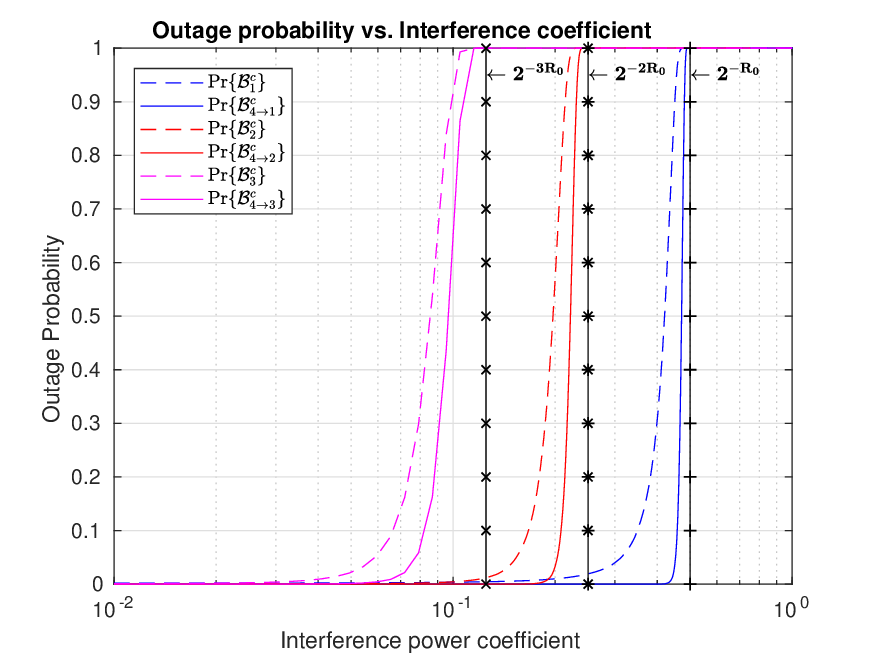}
	\caption{\label{fig:sureout}Certain outage when interference exceeds limits}
\end{figure}

For all of the simulations, there are $K=4$ total users, the number of transmit antennas $M=1$, the number of receive antennas $N=4$, $\beta=1$, and $\xi=3$. Figure \ref{fig:sureout} demonstrates the validity of theorem \ref{thm:power_limit}. The probability of outage for the user $n$ decoding its own signal, and user $K$ decoding the signal for user $n$ are shown together to demonstrate that when interference coefficient $A_n > 2^{-nR_0}$, the outage is certain with probability equal to 1. In the cases of user $n=2$ and $3$, the available total power to users decoding the signal is assumed to be almost $1-2^{(n-1)R_0}$ in order to demonstrate the best case scenario that  every user $l=n,\ldots,K$ could perform SIC on the signal for user $n-1$ with the least amount of power allocated without being in certain outage. 

\begin{figure}
	\centering
	\includegraphics[scale=0.65]{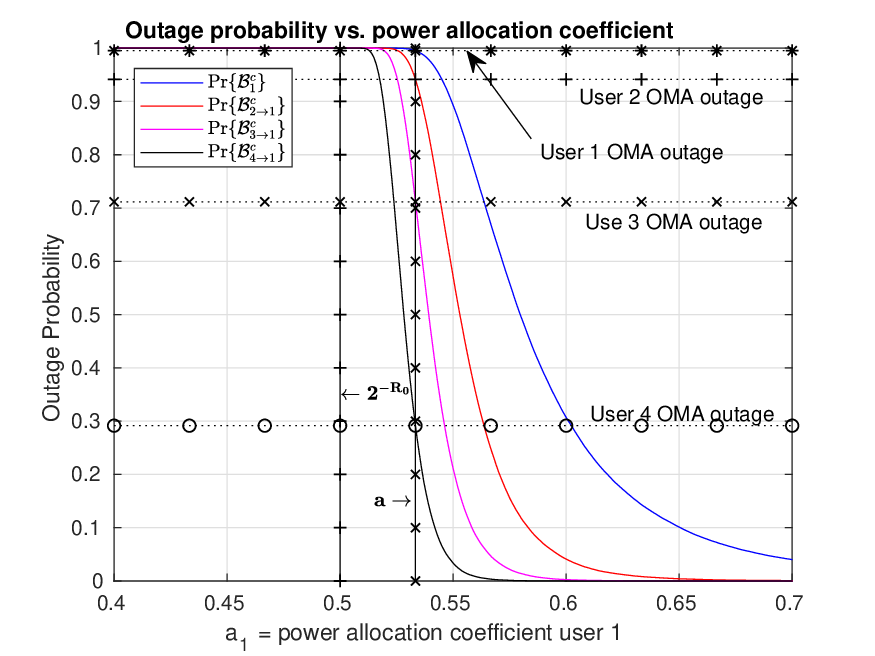}
	\caption{\label{fig:omanoma}Demonstration of lemma \ref{lem:ampower} and theorem \ref{thm:nomaomapower}.}
\end{figure}

Figure \ref{fig:omanoma} demonstrates that the power allocation from lemma \ref{lem:ampower} for the case of user 1, does indeed give the same outage performance as OMA, according to theorem \ref{thm:nomaomapower}. The value of $a$ pointing to the vertical line in the plot refers to $a_1=\frac{2^{3R_0}(2^{R_0}-1)}{2^{4R_0}-1}$, according to lemma \ref{lem:ampower}. Note that user-$1$ has an outage probability nearly equal to 1, which is confirmed for NOMA at $a$, while this value for $a$ is the crossing point for NOMA when users $2-4$ also have the same outage performance as OMA. This plot also shows that the power for user-$1$ can also be increased beyond the value given by lemma \ref{lem:ampower}, which can decrease the probability of outage for this user. Of course, this would be at the expense of the performance of the remaining users rates.

\section{Conclusion and Future Work}\label{sec:conclusion}
It was shown that the fundamental limits of a NOMA system's power allocation coefficients depend solely on the minimum QoS targeted rate and the total number of users simultaneouly using the system. A fixed power allocation strategy was derived that makes the outage events equal at each user, for all signals to be decoded, and it was proven that this strategy provides the case where a fixed-power approach will have the same outage event as an OMA system.

This work can be extended by investigating the case where each user has its unique targetted rate. Furthermore, a comparison of this fixed-power bench mark and dynamic power allocation should be investigated, as given for the multi-user case in \cite{FairNOMAFull}, and what implications exist when the dynamic power allocation approach lies in different subsets of the derived fundamental region found in this work. In other words, is it known that the channel is in outage if a dynamic power allocation coefficient is found to be outside of the region derived in this work?

%%%%%%%%%%%%%%%%%%%%%%%%%%%%%%%%%%%%%%%%%%%%%%%%%%%%%%%
%%%%%%%%%%%%%%%%%%%%%%%%%%%%%%%%%%%%%%%%%%%%%%%%%%%%%%%
% 					BIBLIOGRAPHY	
%%%%%%%%%%%%%%%%%%%%%%%%%%%%%%%%%%%%%%%%%%%%%%%%%%%%%%%
%%%%%%%%%%%%%%%%%%%%%%%%%%%%%%%%%%%%%%%%%%%%%%%%%%%%%%%

\end{document}